\documentclass{ws-ijmpa}
\usepackage[super,compress]{cite}
\usepackage{graphicx}
\begin{document}
\markboth{E. R. Bezerra de Mello,   A. A. Saharian \& A. Mohammadi}
{Induced fermionic current at finite temperature}

%
\catchline{}{}{}{}{}
%

\title{Induced fermionic current by a magnetic flux in a cosmic string spacetime
at finite temperature}

\author{Eug\^enio R. Bezerra de Mello\footnote{emello@fisica.ufpb.br}}

\address{Departamento de F\'{\i}sica-CCEN, Universidade Federal da Para\'{\i}ba\\
J. Pessoa, PB, 58.059-970, Brazil}

\author{Aram A. Saharian\footnote{saharian@ysu.am}}

\address{Department of Physics, Yerevan State University\\
1 Alex Manoogian Street, 0025 Yerevan, Armenia}

\author{Azadeh Mohammadi\footnote{a.mohammadi@fisica.ufpb.br}}

\address{Departamento de F\'{\i}sica-CCEN, Universidade Federal da Para\'{\i}ba\\
J. Pessoa, PB, 58.059-970, Brazil}

\maketitle

\begin{history}
\received{Day Month Year}
\revised{Day Month Year}
\end{history}

\begin{abstract}
Here we analyze the finite temperature expectation values of the charge and
current densities for a massive fermionic quantum field with nonzero chemical
potential, $\mu$, induced by a magnetic flux running along the axis of an idealized cosmic string.
These densities are decomposed into the vacuum expectation values and contributions coming from the particles
and antiparticles.  Specifically the charge density is an even periodic function of the
magnetic flux with the period equal to the quantum flux and an odd function
of the chemical potential.  The only nonzero component of the current density
corresponds to the azimuthal current and it  is an odd periodic function
of the magnetic flux and an even function of the chemical potential.
Both analyzed are developed for the cases where
$|\mu |$ is smaller than the mass of the field quanta, $m$.
\end{abstract}

\ccode{PACS numbers:  03.70.+k, 11.10.Wx, 11.27.+d, 98.80.Cq}

\section{Introduction}
Cosmic strings are among the most important class of linear topological
defects with the conical geometry outside the core \cite{Vile94}. The
formation of these type of topologically stable structures during the
cosmological expansion is predicted in most interesting models of high
energy physics. They have a number of interesting observable consequences,
the detection of which would provide an important link between cosmology and
particle physics.

In quantum field theory, the conical topology of the spacetime due to the presence
of a cosmic string causes a number of interesting physical effects. In
particular, many authors have considered the vacuum polarization effects for scalar,
fermionic and vector fields induced by a planar angle deficit.
In addition to the deficit angle parameter, the physical origin of a cosmic
string is characterized by the gauge field flux parameter describing a
magnetic flux running along the string's core. The latter induces additional
polarization effects for charged fields. \cite{Dowk87}-\cite{Site12}  Though
the gauge field strength vanishes outside the string's core, the nonvanishing
vector potential leads to Aharonov-Bohm-like effects on scattering cross
sections and on particle production rates around the cosmic string. \cite{Alfo89}.
For charged fields, the magnetic flux along the string core induces nonzero
vacuum expectation value of the current density. The latter, in addition to
the expectation values of the field squared and the energy-momentum tensor,
is among the most important local characteristics of the vacuum state for
quantum fields. The azimuthal current density for scalar and fermionic fields, induced by a
magnetic flux in the geometry of a straight cosmic string, has been
investigated in  Ref.s~\refcite{Srir01}-\refcite{Brag14}.

Here we shall  consider
the effects of the finite temperature and nonzero chemical potential on the
expectation values of the charge and current densities for a massive
fermionic field in the geometry of a straight cosmic string for arbitrary
values of the planar angle deficit.

\section{Geometry and Fermionic Modes}
The background geometry corresponding to a straight cosmic string lying
along the $z$-axis can be written through the line element
\begin{equation}
ds^{2}=dt^{2}-dr^{2}-r^{2}d\phi ^{2}-dz{}^{2}\ ,  \label{ds21}
\end{equation}%
where $r\geqslant 0$, $0\leqslant \phi \leqslant \phi _{0}=2\pi /q$, $%
-\infty <t<+\infty $. The parameter $q\geqslant 1$ codifies the planar angle
deficit.  In the presence of an external electromagnetic
field with the vector potential $A_{\mu }$, the dynamics of a massive
charged spinor field in curved spacetime is described by the Dirac equation,
\begin{equation}
(i\gamma ^{\mu }{\mathcal{D}}_{\mu }-m)\psi =0\ ,\ {\mathcal{D}}_{\mu
}=\partial _{\mu }+\Gamma _{\mu }+ieA_{\mu },  \label{Direq}
\end{equation}%
where $\gamma ^{\mu }$ are the Dirac matrices in curved spacetime and $%
\Gamma _{\mu }$ are the spin connections. For the geometry at hand the gamma
matrices can be taken in the form
\begin{equation}
\gamma ^{0}=\left(
\begin{array}{cc}
1 & 0 \\
0 & -1%
\end{array}%
\right) ,\;\gamma ^{l}=\left(
\begin{array}{cc}
0 & \rho ^{l} \\
-\rho ^{l} & 0%
\end{array}%
\right) ,  \label{gamcurved}
\end{equation}%
where the $2\times 2$ matrices $\rho ^{l}$ are
\begin{equation}
\rho ^{1}=\left(
\begin{array}{cc}
0 & e^{-iq\phi } \\
e^{iq\phi } & 0%
\end{array}%
\right) \ ,\ \rho ^{2}=-\frac{i}{r}\left(
\begin{array}{cc}
0 & e^{-iq\phi } \\
-e^{iq\phi } & 0%
\end{array}%
\right) \ ,\ \rho ^{3}=\left(
\begin{array}{cc}
1 & 0 \\
0 & -1%
\end{array}%
\right) \ .  \label{betl}
\end{equation}

We shall admit the existence of a gauge field with a constant vector
potential as
\begin{equation}
A_{\mu }=(0,0,A_{\phi},0)\ .  \label{Amu}
\end{equation}%
The azimuthal component $A_{\phi}$ is related to an infinitesimal thin magnetic
flux, $\Phi $, running along the string by $A_{\phi}=-q\Phi /(2\pi )$.

The field operator can be expanded in term of a complete set
of normalized positive- and negative-energy solution
 $\{\psi _{\sigma }^{(+)},\psi _{\sigma }^{(-)}\}$
 of (\ref{Direq}), specified by a set of quantum numbers $\sigma$,  as:
\begin{equation}
\psi =\sum_{\sigma }[\hat{a}_{\sigma }\psi _{\sigma }^{(+)}+\hat{b}_{\sigma
}^{+}\psi _{\sigma }^{(-)}]  \ ,  \label{psiexp}
\end{equation}
where $\hat{a}_{\sigma }$ and $\hat{b}_{\sigma }^{+}$ represent the
annihilation and creation operators corresponding to particles and
antiparticles respectively.

Here, we are interested in the effects of the presence of the cosmic string
and magnetic flux on the expectation values of the charge and current
densities assuming that the field is in thermal equilibrium at finite
temperature $T$.  The standard form
of the density matrix for the thermodynamical equilibrium distribution at
temperature $T$ is
\begin{equation}
\hat{\rho}=Z^{-1}e^{-\beta (\hat{H}-\mu ^{\prime }\hat{Q})},\;\beta =1/T \ \  , \ {\rm and} \
Z=\mathrm{tr}[e^{-\beta (\hat{H}-\mu ^{\prime }\hat{Q})}] \   ,
\label{rho}
\end{equation}
where $\hat{H}$ is the Hamilton operator, $\hat{Q}$ denotes a conserved
charge and $\mu ^{\prime }$ is the corresponding chemical potential.
The thermal average of the creation and annihilation operators are given by:
\begin{eqnarray}
\mathrm{tr}[\hat{\rho}\hat{a}_{\sigma }^{+}\hat{a}_{\sigma ^{\prime }}] &=&
\frac{\delta _{\sigma \sigma ^{\prime }}}{e^{\beta (\varepsilon _{\sigma
}^{(+)}-\mu )}+1},  \notag \\
\mathrm{tr}[\hat{\rho}\hat{b}_{\sigma }^{+}\hat{b}_{\sigma ^{\prime }}] &=&
\frac{\delta _{\sigma \sigma ^{\prime }}}{e^{\beta (\varepsilon _{\sigma
}^{(-)}+\mu )}+1},  \label{traa}
\end{eqnarray}%
where $\mu =e\mu ^{\prime }$ and $\pm \varepsilon _{\sigma }^{(\pm )}$ with $
\varepsilon _{\sigma }^{(\pm )}>0$, are the energies corresponding to the
modes $\psi _{\sigma }^{(\pm )}$.

The expectation value of the fermionic current density  given by
$\left\langle j^{\nu }\right\rangle =e\,\mathrm{tr}[\hat{\rho}\bar{\psi}
(x)\gamma ^{\nu }\psi (x)]$, can be expressed by
\begin{equation}
\left\langle j^{\nu }\right\rangle =\left\langle j^{\nu }\right\rangle
_{0}+\sum_{\chi =+,-}\left\langle j^{\nu }\right\rangle _{\chi },  \label{C1}
\end{equation}%
where
\begin{equation}
\left\langle j^{\nu }\right\rangle _{0}=e\sum_{\sigma }\bar{\psi}_{\sigma
}^{(-)}(x)\gamma ^{\nu }\psi _{\sigma }^{(-)}(x),  \label{Cvev}
\end{equation}%
is the vacuum expectation value and
\begin{equation}
\left\langle j^{\nu }\right\rangle _{\pm }=\pm e\sum_{\sigma }\frac{\bar{\psi
}_{\sigma }^{(\pm )}\gamma ^{\nu }\psi _{\sigma }^{(\pm )}}{e^{\beta
(\varepsilon _{\sigma }^{(\pm )}\mp \mu )}+1}.  \label{jpm}
\end{equation}%
Here, $\left\langle j^{\nu }\right\rangle _{\pm }$ is the part in the
expectation value coming from the particles for the upper sign and from the
antiparticles for the lower sign.

We shall use the normalized  fermionic modes found in  Ref.~\refcite{Beze13}
specified by the set of quantum numbers $\sigma =(\lambda ,k,j,s)$ with
\begin{equation}
\lambda \geqslant 0,\;-\infty <k<+\infty ,\;j=\pm 1/2,\pm 3/2,\ldots
,\;s=\pm 1.  \label{range}
\end{equation}
These functions are expressed as
\begin{equation}
\psi _{\sigma }^{(\pm )}(x)=C_{\sigma }^{(\pm )}e^{\mp iEt+ikz+iqj\phi
}\left(
\begin{array}{c}
J_{\beta _{j}}(\lambda r)e^{-iq\phi /2} \\
sJ_{\beta _{j}+\epsilon _{j}}(\lambda r)e^{iq\phi /2} \\
\pm \frac{k-is\epsilon _{j}\lambda }{E\pm m}J_{\beta _{j}}(\lambda
r)e^{-iq\phi /2} \\
\mp s\frac{k-is\lambda \epsilon _{j}}{E\pm m}J_{\beta _{j}+\epsilon
_{j}}(\lambda r)e^{iq\phi /2}%
\end{array}%
\right) \ ,  \label{psi+n}
\end{equation}%
where $J_{\nu }(x)$ is the Bessel function, $|C_{\sigma }^{(\pm )}|^{2}=\frac{q\lambda (E\pm m)}{16\pi ^{2}E}$ and
\begin{eqnarray}
E=\varepsilon _{\sigma }^{(\pm )}=\sqrt{\lambda ^{2}+k^{2}+m^{2}} \  ,  \
\beta _{j}=q|j+\alpha |-\epsilon _{j}/2\ ,\;\alpha =eA_{\phi}/q=-\Phi /\Phi
_{0}  \  ,
\end{eqnarray}
with $\epsilon _{j}=\mathrm{sgn}(j+\alpha )$ and $\Phi _{0}=2\pi /e$ being
the flux quantum.

\section{Charge Density}
We start with the charge density corresponding to the $\nu =0$ component of (\ref{C1}).
In Ref.~\refcite{Beze13} we have explicitly shown that the formal expression for the vacuum expectation value  of 
charge density is given in terms of a divergent integral. In order to obtain a finite and well defined
result we introduced a cutoff function. With this cutoff the integral could be evaluated. 
Our next steps were to subtract the Minkowskian part $(\alpha_0=0, \ q=1)$ and to
remove the cutoff function. As final result a vanishing value for the renormalized charge density
was obtained. 

Substituting the mode functions (\ref{psi+n}) into (\ref{jpm}),
for the contributions coming from the particles and antiparticles we get
\begin{equation}
\left\langle j^{0}\right\rangle _{\pm }=\pm \frac{eq}{8\pi ^{2}}\sum_{\sigma
}\lambda \frac{J_{\beta _{j}}^{2}(\lambda r)+J_{\beta _{j}+\epsilon
_{j}}^{2}(\lambda r)}{e^{\beta (E\mp \mu )}+1},  \label{j0pm}
\end{equation}
where we use the notation
\begin{equation}
\sum_{\sigma }=\int_{-\infty }^{+\infty }dk\int_{0}^{\infty }d\lambda \
\sum_{s=\pm 1}\sum_{j}\ .  \label{Sumsig}
\end{equation}
In the case $\mu =0$ the contributions from the particles and
antiparticles cancel each other and the total charge density,%
\begin{equation}
\left\langle j^{0}\right\rangle =\left\langle j^{0}\right\rangle
_{+}+\left\langle j^{0}\right\rangle _{-},  \label{j0tot}
\end{equation}%
is zero.  From (\ref{j0pm}) one can see that the charge density is an even periodic
function of the parameter $\alpha $ with the period equal to 1. Consequently
the charge density is a periodic function of the magnetic flux with the
period equal to the quantum flux. If we present this parameter as
\begin{equation}
\alpha =n_{0}+\alpha _{0},\;|\alpha _{0}|\leqslant 1/2,  \label{alf0}
\end{equation}%
with $n_{0}$ being an integer, then the current density depends on $\alpha
_{0}$ alone.

Here in this paper we shall consider only  the case $|\mu |<m$.\footnote{The analysis
for the case  $|\mu |>m$ is given in  Ref.~\refcite{Mello}.}  By using the expansion
$(e^{y}+1)^{-1}=-\sum_{n=1}^{\infty }(-1)^{n}e^{-ny}$,
 the charge densities for particles and antiparticles can be
presented in the form
\begin{eqnarray}
\left\langle j^{0}\right\rangle _{\pm } &=&\mp \frac{eq\beta }{4\pi ^{2}r^{4}%
}\ \sum_{n=1}^{\infty }(-1)^{n}ne^{\pm n\beta \mu }\int_{0}^{\infty }dx\,x
\notag \\
&&\times F(q,\alpha _{0},x)e^{-m^{2}r^{2}/2x-(1+n^{2}\beta ^{2}/2r^{2})x}\ ,
\label{j0pm2}
\end{eqnarray}%
where the notation%
\begin{equation}
F(q,\alpha _{0},x)=\sum_{j}\ \left[ I_{\beta _{j}}(x)+I_{\beta _{j}+\epsilon
_{j}}(x)\right] ,  \label{Fq}
\end{equation}
is introduced. In  Ref.~\refcite{Beze10b}   we have shown that
\begin{eqnarray}
F(q,\alpha _{0},x) &=&\frac{4}{q}\left[ \frac{e^{x}}{2}%
+\sum_{k=1}^{[q/2]}(-1)^{k}c_{k}\cos \left( 2\pi k\alpha _{0}\right)
e^{x\cos (2\pi k/q)}\right.  \notag \\
&&\left. +\frac{q}{\pi }\int_{0}^{\infty }dy\frac{h(q,\alpha _{0},2y)\sinh y%
}{\cosh (2qy)-\cos (q\pi )}e^{-x\cosh {(2y)}}\right] \ ,  \label{Fq1}
\end{eqnarray}%
where $[q/2]$ means the integer part of $q/2$ and the notation%
\begin{equation}
h(q,\alpha _{0},x)=\sum_{\chi =\pm 1}\cos \left[ \left( 1/2+\chi \alpha
_{0}\right) q\pi \right] \sinh \left[ \left( 1/2-\chi \alpha _{0}\right) qx%
\right] ,  \label{h}
\end{equation}%
is assumed. Here and in what follows we use the notations%
\begin{equation}
c_{k}=\cos {(\pi k/q),\;s}_{k}=\sin {(\pi k/q).}  \label{cksk}
\end{equation}
Substituting (\ref{Fq1}) into (\ref{j0pm2}), after integration over $x$, we
find the expression%
\begin{eqnarray}
\left\langle j^{0}\right\rangle _{\pm } &=&\left\langle j^{0}\right\rangle _{%
\mathrm{M}\pm }\mp \frac{2em^{4}\beta }{\pi ^{2}}\ \sum_{n=1}^{\infty
}(-1)^{n}ne^{\pm n\beta \mu }  \notag \\
&&\times \left[ \sum_{k=1}^{[q/2]}(-1)^{k}c_{k}\cos \left( 2\pi k\alpha
_{0}\right) f_{2}(m\beta s_{n}(r/\beta ,k/q))\right.  \notag \\
&&\left. +\frac{q}{\pi }\int_{0}^{\infty }dy\frac{\sinh \left( y\right)
h(q,\alpha _{0},2y)}{\cosh (2qy)-\cos (q\pi )}f_{2}(m\beta c_{n}(r/\beta ,y))%
\right] ,  \label{j0pm3}
\end{eqnarray}%
where%
\begin{equation}
\left\langle j^{0}\right\rangle _{\mathrm{M}\pm }=\mp \frac{e\beta m^{4}}{%
\pi ^{2}}\sum_{n=1}^{\infty }(-1)^{n}ne^{\pm n\beta \mu }f_{2}(nm\beta ),
\label{j0pmM}
\end{equation}%
is the corresponding charge density in Minkowski spacetime in the absence of
the magnetic flux and the cosmic string ($\alpha _{0}=0$, $q=1$). Here we
have introduced the notations%
\begin{equation}
f_{\nu }(x)=x^{-\nu }K_{\nu }(x),  \label{fnu1}
\end{equation}%
with $K_{\nu }(x)$ being the MacDonald function and
\begin{eqnarray}
s_{n}(x,y) =\sqrt{n^{2}+4x^{2}\sin ^{2}(\pi y)}, \
c_{n}(x,y) =\sqrt{n^{2}+4x^{2}\cosh ^{2}y}\ .  \label{sncn}
\end{eqnarray}

\begin{figure}[tbph]
{\includegraphics[width=6.0cm]{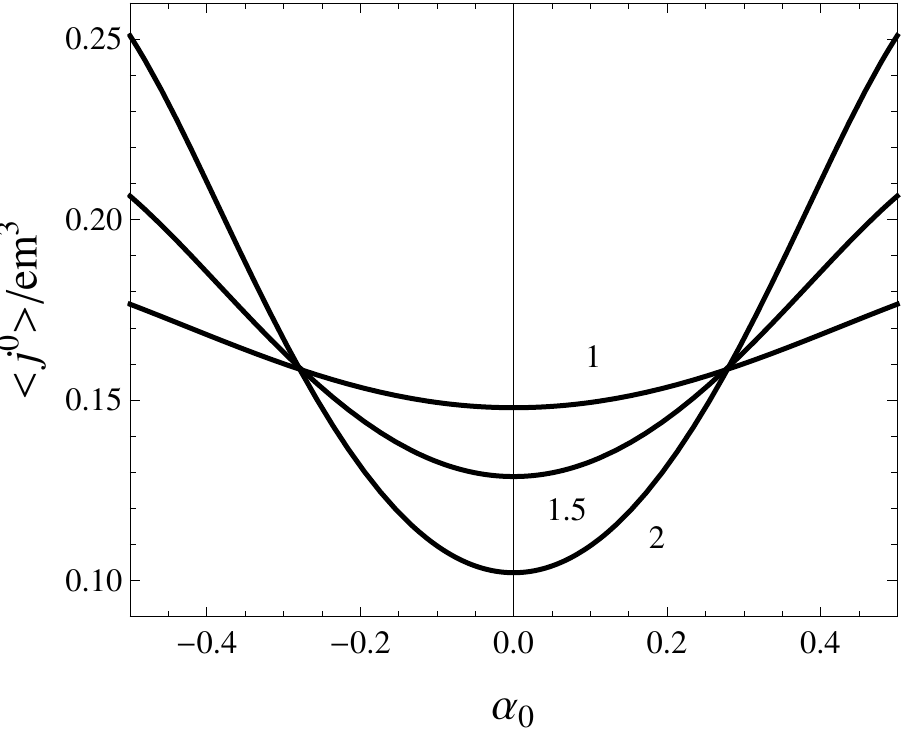}}
{\includegraphics[width=6.2cm]{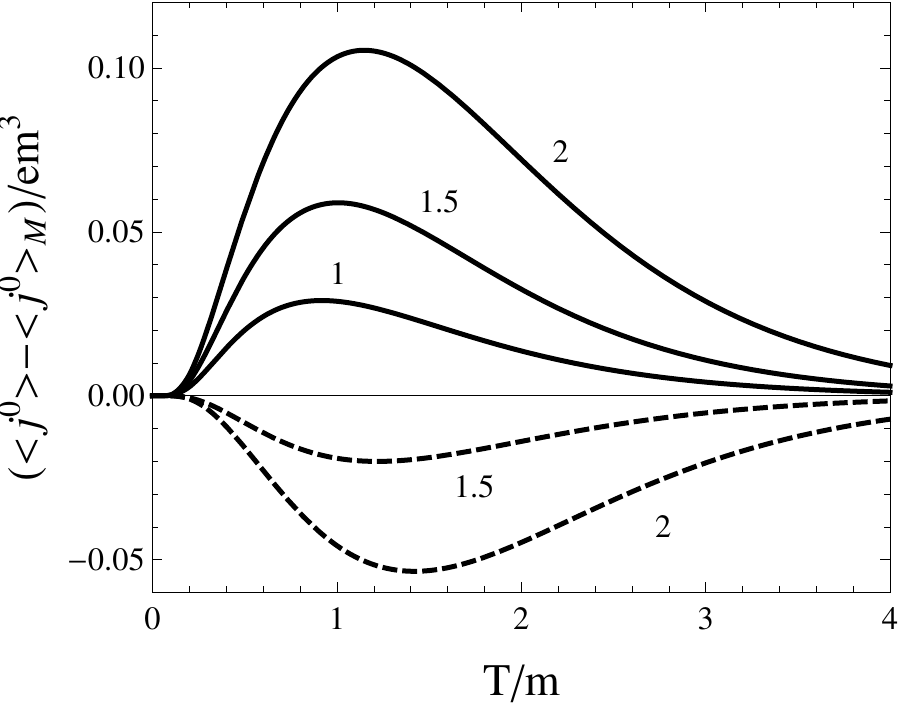}}
\caption{The total charge density as a function of the parameter $\protect
\alpha _{0}$ (left panel) and the charge density induced by the string and
magnetic flux as a function of the temperature (right panel). The numbers
near the curves correspond to the values of the parameter $q$. The graphs on
the left panel are plotted for $T/m=1$, $mr=0.25 $, and $\protect\mu /m=0.5$
. On the right panel, the full and dashed curves correspond to the values $
\protect\alpha _{0}=1/2$ and $\protect\alpha _{0}=0$, respectively. For the
graphs on the right panel we have taken $\protect\mu /m=0.5$ and $mr=1/8$ \label{fig1}}
\end{figure}

For the total charge density one gets
\begin{eqnarray}
\left\langle j^{0}\right\rangle &=&-\frac{4em^{4}\beta }{\pi ^{2}}\
\sum_{n=1}^{\infty }(-1)^{n}n\sinh (n\beta \mu )\left[ \frac{1}{2}
f_{2}(m\beta n)\right.  \notag \\
&&+\sum_{k=1}^{[q/2]}(-1)^{k}c_{k}\cos \left( 2\pi k\alpha _{0}\right)
f_{2}(m\beta s_{n}(r/\beta ,k/q))  \notag \\
&&\left. +\frac{q}{\pi }\int_{0}^{\infty }dy\frac{h(q,\alpha _{0},2y)\sinh y
}{\cosh (2qy)-\cos (q\pi )}f_{2}(m\beta c_{n}(r/\beta ,y))\right]  \  .
\label{j0}
\end{eqnarray}
We present in figure \ref{fig1}, the total charge density as a function
of the parameter $\alpha _{0}$ (left panel) and the charge density induced
by the string and magnetic flux as a function of the temperature (right
panel). The numbers near the curves correspond to the values of the
parameter $q$. The graphs on the left panel are plotted for $T/m=1$, $
mr=0.25 $, and $\mu /m=0.5$. On the right panel, the full and dashed curves
correspond to the values $\alpha _{0}=1/2$ and $\alpha _{0}=0$, respectively
(note that for $q=1$, $\alpha _{0}=0$ one has $\left\langle
j^{0}\right\rangle -\left\langle j^{0}\right\rangle _{\mathrm{M}}=0$). For
the graphs on the right panel we have taken $\mu /m=0.5$ and $mr=1/8$.

\begin{figure}[tbph]
\begin{center}
{\includegraphics[width=6.0cm]{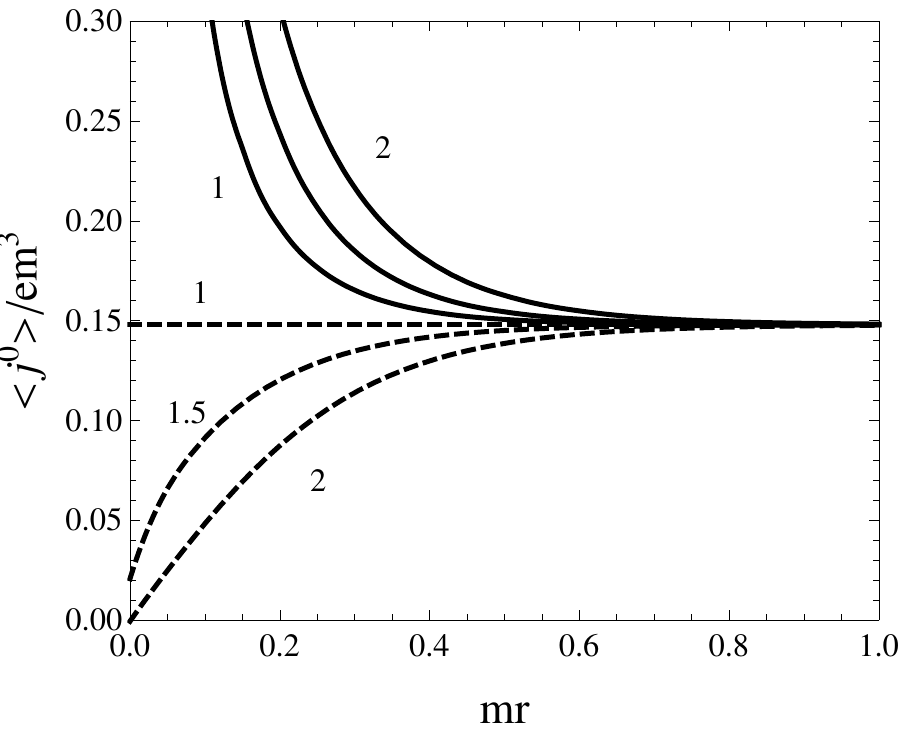}}
\caption{The charge density versus the radial coordinate. The full and
dashed lines correspond to the values $\protect\alpha _{0}=1/2$ and $\protect%
\alpha _{0}=0$, respectively. The numbers near the curves present the values
of the parameter $q$. The graphs are plotted for $T/m=1$
and $\protect\mu /m=0.5$.}
\label{fig2}
\end{center}
\end{figure}

At large distance and high temperature the Minkowski contribution
dominates and the one induced by the cosmic string and magnetic flux  are exponentially
suppressed.\cite{Mello}.
In figure \ref{fig2}, we present the charge density as a function of the
radial coordinate. The full and dashed lines correspond to the values $
\alpha _{0}=1/2$ and $\alpha _{0}=0$, respectively. The numbers near the
curves present the values of the parameter $q$. The graphs
are plotted for $T/m=1$ and $\mu /m=0.5$.

\section{Azimuthal Current}

Now we turn to the investigation of the current density. The only nonzero
component corresponds to the azimuthal current ($\nu =2$ in (\ref{C1})). By
taking into account the expression for the mode functions, from (\ref{jpm})
for the physical components of the current densities of the particles and
antiparticles, $\left\langle j_{\phi }\right\rangle _{\pm }=r\left\langle
j^{2}\right\rangle _{\pm }$, we get%
\begin{equation}
\left\langle j_{\phi }\right\rangle _{\pm }=\frac{eq}{4\pi ^{2}}\sum_{\sigma
}\epsilon _{j}\frac{\lambda ^{2}}{E}\frac{J_{\beta _{j}}(\lambda r)J_{\beta
_{j}+\epsilon _{j}}(\lambda r)}{e^{\beta (E\mp \mu )}+1},  \label{j2pm}
\end{equation}%
where the upper and lower signs correspond to the particles and
antiparticles respectively and the collective summation is defined by (\ref%
{Sumsig}).

For the case $|\mu |<m$,  by using the same expansion  for the
denominator as we did in the previous case, the current densities (\ref{j2pm})
reads
\begin{equation}
\left\langle j_{\phi }\right\rangle _{\pm }=-\frac{eq}{2\pi ^{2}r^{3}}
\sum_{n=1}^{\infty }(-1)^{n}e^{\mp n\beta \mu }\int_{0}^{\infty
}dx\,xe^{-m^{2}r^{2}/(2x)-(1+n^{2}\beta ^{2}/(2r^{2}))x}G(q,\alpha _{0},x),
\label{j2pm3}
\end{equation}
with the notation
\begin{equation}
G(q,\alpha _{0},x)=\sum_{j}\ \left[ I_{\beta _{j}}(x)-I_{\beta _{j}+\epsilon
_{j}}(x)\right] .  \label{Gq}
\end{equation}

By using the integral representation for the modified Bessel function we can write
\begin{eqnarray}
G(q,\alpha _{0},x) &=&\frac{4}{q}\sideset{}{'}{\sum}
_{k=1}^{[q/2]}(-1)^{k}s_{k}\sin \left( 2\pi k\alpha _{0}\right) e^{x\cos
(2\pi k/q)}  \notag \\
&&+\frac{4}{\pi }\int_{0}^{\infty }dy\ \frac{g(q,\alpha _{0},2y)\cosh y}{
\cosh (2qy)-\cos (q\pi )}e^{-x\cosh {(2y)}},  \label{Gq1}
\end{eqnarray}
with the notation
\begin{equation}
g(q,\alpha _{0},x)=\sum_{\chi =\pm 1}\chi \cos \left[ \left( 1/2+\chi \alpha
_{0}\right) q\pi \right] \cosh \left[ \left( 1/2-\chi \alpha _{0}\right) qx
\right] \ .  \label{g}
\end{equation}
The prime on the summation sign in (\ref{Gq1}) means that, in the case where
$q$ is an even number, the term with $k=q/2$ should be taken with the
coefficient 1/2.

Substituting (\ref{Gq1}) into (\ref{j2pm3}), after integrating over $x$, we
obtain
\begin{align}
\left\langle j_{\phi }\right\rangle _{\pm }& =-\frac{4em^{4}r}{\pi ^{2}}
\sum_{n=1}^{\infty }(-1)^{n}e^{\mp n\beta \mu }  \notag \\
& \times \left[ \ \sideset{}{'}{\sum}_{k=1}^{[q/2]}(-1)^{k}s_{k}\sin \left(
2\pi k\alpha _{0}\right) f_{2}\left( m\beta s_{n}(r/\beta ,k/q)\right)
\right.  \notag \\
& +\left. \frac{q}{\pi }\int_{0}^{\infty }dy\ \frac{\cosh \left( y\right)
g(q,\alpha _{0},2y)}{\cosh (2qy)-\cos (q\pi )}f_{2}(m\beta c_{n}(r/\beta ,y))
\right] ,  \label{j2pm4}
\end{align}
where the functions in the arguments of $f_{2}(x)$ are defined by (\ref{sncn}).
For the case $q=1$, i.e. in the absence of conical defect, the above expression reduces
to\footnote{The Minkowskian contribution $(q=1, \ \alpha_0=0)$ vanishes.}
\begin{align}
\left\langle j_{\phi }\right\rangle _{\pm }& =\frac{4em^{4}r}{\pi ^{3}}\sin
(\alpha _{0}\pi )\sum_{n=1}^{\infty }(-1)^{n}e^{\mp n\beta \mu }  \notag \\
& \times \int_{0}^{\infty }dy\,\cosh (2\alpha _{0}y)f_{2}\left( m\beta
c_{n}(r/\beta ,y)\right) \  .  \label{j2pmq1}
\end{align}

Taking into account the expression for the vacuum expectation value
of the current density from \cite{Beze13}, the total current density reads
\begin{align}
\left\langle j_{\phi }\right\rangle & =-\frac{8em^{4}r}{\pi ^{2}}%
\sideset{}{'}{\sum}_{n=0}^{\infty }(-1)^{n}\cosh (n\beta \mu )  \notag \\
& \times \left[ \ \sideset{}{'}{\sum}_{k=1}^{[q/2]}(-1)^{k}s_{k}\sin \left(
2\pi k\alpha _{0}\right) f_{2}\left( m\beta s_{n}(r/\beta ,k/q)\right)
\right.  \notag \\
& +\left. \frac{q}{\pi }\int_{0}^{\infty }dy\ \frac{\cosh \left( y\right)
g(q,\alpha _{0},2y)}{\cosh (2qy)-\cos (q\pi )}f_{2}\left( m\beta
c_{n}(r/\beta ,y)\right) \right] ,  \label{j2}
\end{align}%
where the prime on the sign of the summation over $n$ means that the term $
n=0$ should be taken with the coefficient 1/2. This term corresponds to the
vacuum expectation value of the current density, $\left\langle j_{\phi
}\right\rangle _{0}$.

Now we would like to analyze the case of a massless field. Because of the condition $|\mu |\leqslant m$, we
should also take $\mu =0$. By using the asymptotic expression for the
MacDonald function for small  argument, the summation over $n$
takes the form $\sideset{}{'}{\sum}_{n=0}^{\infty
}(-1)^{n}\left( n^{2}+x^{2}\right) ^{-2}$, that can be
expressed in terms of the hyperbolic functions. So we get
\begin{align}
\left\langle j_{\phi }\right\rangle & =-\frac{eT}{2\pi r^{2}}\left[ \ %
\sideset{}{'}{\sum}_{k=1}^{[q/2]}\frac{(-1)^{k}}{s_{k}^{2}}\sin \left( 2\pi
k\alpha _{0}\right) h(2rTs_{k})\right.  \notag \\
& \left. +\frac{q}{\pi }\int_{0}^{\infty }dy\ \frac{g(q,\alpha _{0},2y)}{%
\cosh (2qy)-\cos (q\pi )}\frac{h(2rT\cosh y)}{\cosh ^{2}y}\right] \ ,
\label{j2pmm0}
\end{align}%
where we have introduced the function%
\begin{equation}
h(x)=\frac{1+\pi x\coth (\pi x)}{\sinh (\pi x)}.  \label{hx}
\end{equation}

\begin{figure}[tbph]
{\includegraphics[width=6.0cm]{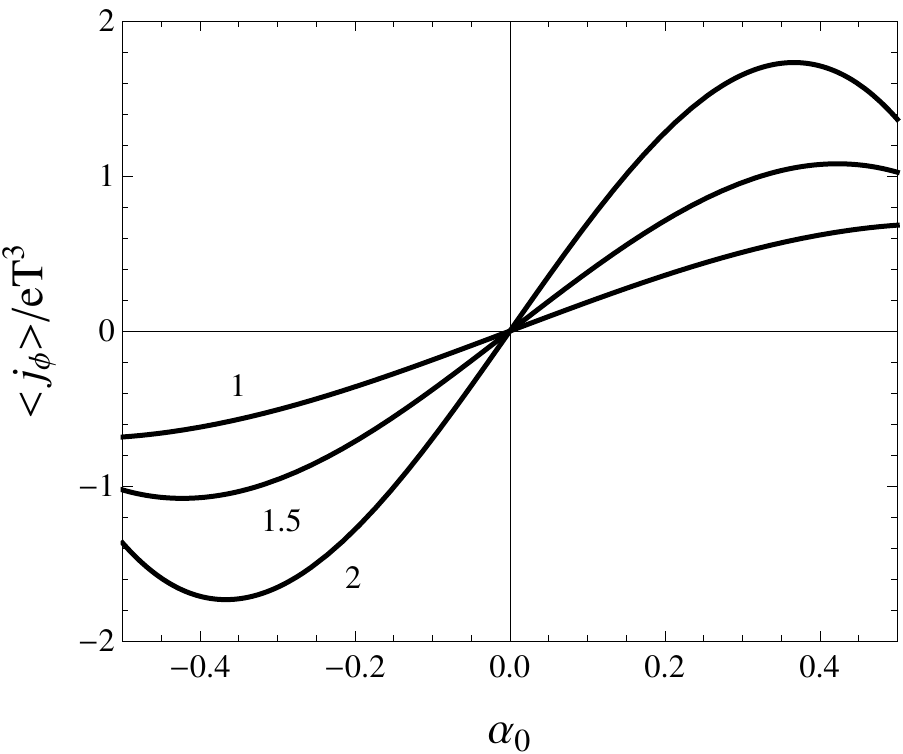}}
{\includegraphics[width=6.2cm]{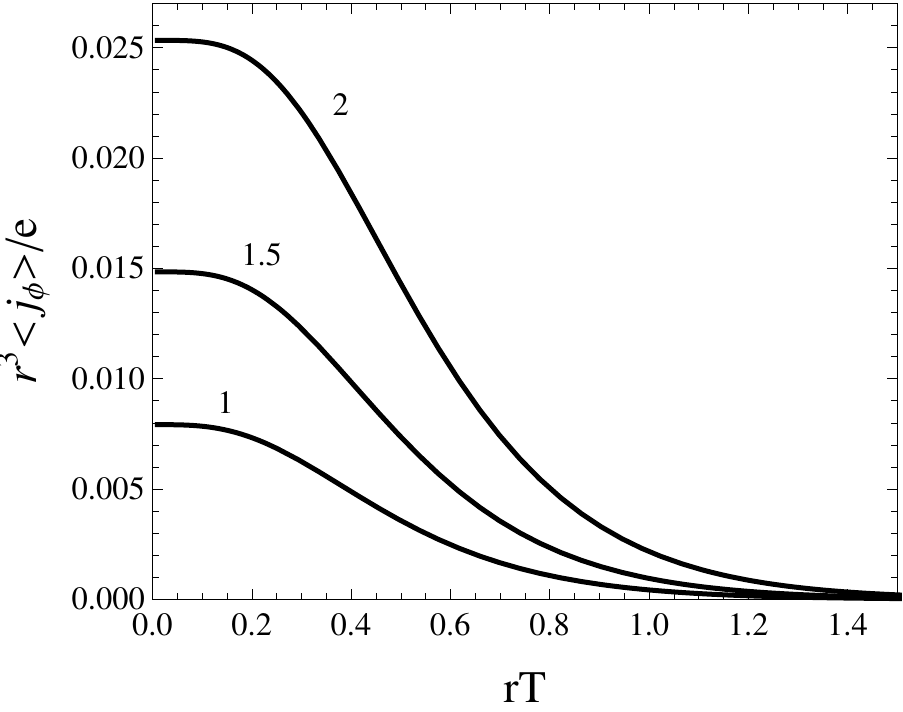}}
\caption{The azimuthal current density for a massless field with zero
chemical potential as a function of $\protect\alpha _{0}$ (left panel) and
as a function of the temperature (right panel). The numbers near the curves
correspond to the values of $q$. For the graphs on the left panel $rT=0.25$
and for the right panel $\protect\alpha _{0}=0.25$.\label{fig3}}
\end{figure}

We plot in  figure \ref{fig3}, for a massless field with $\mu =0$,
the azimuthal current density as a function of the parameter $\alpha _{0}$
(left panel) and as a function of the temperature (right panel). The numbers
near the curves correspond to the values of $q$. In the graphs on the left
panel we assume $rT=0.25$ and on the right  $\alpha _{0}=0.25$.

\section{Conclusion}

\label{sec:Conc}

In this paper, we have analyzed the combined effects of the planar angle
deficit and the magnetic flux on the charge and current densities for a
massive fermionic field at thermal equilibrium  considering nonzero chemical potential.
These densities are decomposed into the vacuum expectation values and finite
temperature contributions, coming from the particles and antiparticles.

For the charge density the renormalized vacuum expectation value vanishes and the
expectation value for the particles and antiparticles in the case $|\mu
|\leqslant m$ are given by (\ref{j0pm3}). The charge
density is an even periodic function of the magnetic flux with period equal
to the quantum flux. For the zero chemical potential the contributions from
the particles and antiparticles cancel each other and the total charge
density, given by (\ref{j0}), vanishes.

The only nonzero component of the expectation value for the current density
corresponds to the current along the azimuthal direction. This current
vanishes in the absence of the magnetic flux and is an odd periodic function
of the latter with the period equal to the quantum flux. The azimuthal
current density is an even function of the chemical potential. For the zero
chemical potential, the contributions to the total current density from the
particles and antiparticles coincide.

\section*{Acknowledgments}
The authors thank to brazilian agency CNPq  for partial financial support.


\begin{thebibliography}{0}

\bibitem{Vile94} A. Vilenkin and E. P. S. Shellard, \textit{Cosmic Strings and
Other Topological Defects} (Cambridge University Press, Cambridge, 1994).

\bibitem{Dowk87} J. S. Dowker, {\it Phys. Rev. D} \textbf{36}, 3095 (1987).

\bibitem{Dowker87} J. S. Dowker, {\it Phys. Rev. D} \textbf{36}, 3742 (1987).

\bibitem{Guim94} M. E. X. Guimar\~{a}es and B. Linet, {\it Commun. Math. Phys.}
\textbf{165}, 297 (1994).

\bibitem{Spin03} J. Spinelly and E. R. Bezerra de Mello, {\it Class. Quantum Grav.}
\textbf{20} 874, (2003).

\bibitem{Spinelly02} J. Spinelly and E.R. Bezerra de Mello, {\it Int. J. Mod.
Phys. A} \textbf{17}, 4375 (2002).

\bibitem{Spin04} J. Spinelly and E. R. Bezerra de Mello, {\it Int. J. Mod. Phys. D}
\textbf{13}, 607 (2004).

\bibitem{Spin08} J. Spinelly and E. R. Bezerra de Mello, {\it JHEP} \textbf{09},
005 (2008).

\bibitem{Site12} Yu. A. Sitenko and N. D. Vlasii, {\it Class. Quantum Grav.} \textbf{%
29}, 095002 (2012).

\bibitem{Alfo89} M. G. Alford and F. Wilczek, {\it Phys. Rev. Lett.} \textbf{62},
1071 (1989).

\bibitem{Vacha11} D. A. Steer and T. Vachaspati, {\it Phys. Rev. D} \textbf{83}, 043528 (2011).

\bibitem{Srir01} L. Sriramkumar, {\it Class. Quantum Grav.} \textbf{18}, 1015
(2001).

\bibitem{Site09} Yu. A. Sitenko and N. D. Vlasii, {\it Class. Quantum Grav.} \textbf{%
26}, 195009 (2009).

\bibitem{Beze10} E. R. Bezerra de Mello, {\it Class. Quantum Grav.} \textbf{27},
095017 (2010).

\bibitem{Beze13} E. R. Bezerra de Mello and A. A. Saharian, {\it Eur. Phys. J. C}
\textbf{73}, 2532 (2013).

\bibitem{Brag14} E. A. F. Bragan\c{c}a, H. F. Santana Mota and E. R. Bezerra
de Mello, {\it Int. J. Mod. Phys. D} {\bf 24}, 1550055 (2015).

\bibitem{Beze13} E. R. Bezerra de Mello and A. A. Saharian, {\it Eur. Phys. J. C}
\textbf{73}, 2532 (2013).

\bibitem{Mello} A. Mohammadi, E. R. Bezerra de Mello and A. A. Saharian,
{\it J. Phys. A. Math. Theor.} {\bf 48} 185401 (2015).

 \bibitem{Beze10b} E. R. Bezerra de Mello, V. B. Bezerra, A. A. Saharian and
V. M. Bardeghyan, {\it Phys. Rev. D} \textbf{82}, 085033 (2010).

\end{thebibliography}
\end{document}